\documentclass[twocolumn]{aastex631}
\usepackage{amssymb}
\usepackage{array}
\usepackage{multirow}
\usepackage{bold-extra}

\graphicspath{{./}{figures/}}
\AtBeginDocument{
}

\newcommand{\black}{\color{black}}
\newcommand{\red}{\black}

\shorttitle{Galactic neutrinos with Baikal-GVD}
\shortauthors{Baikal-GVD et al.}

\begin{document}

\title{Probing the Galactic neutrino flux at neutrino energies above 200~TeV with the Baikal Gigaton Volume Detector}

\correspondingauthor{S.~V.~Troitsky}
\email{st@ms2.inr.ac.ru}

\author{V.~A.~Allakhverdyan}
\affiliation{Joint Institute for Nuclear Research, Dubna, 141980  Russia}
\author{A.~D.~Avrorin}
\affiliation{Institute for Nuclear
Research of the Russian Academy of Sciences, 60th October Anniversary Prospect 7a, Moscow 117312, Russia}
\author{A.~V.~Avrorin}
\affiliation{Institute for Nuclear
Research of the Russian Academy of Sciences, 60th October Anniversary Prospect 7a, Moscow 117312, Russia}
\author{V.~M.~ Aynutdinov}
\affiliation{Institute for Nuclear
Research of the Russian Academy of Sciences, 60th October Anniversary Prospect 7a, Moscow 117312, Russia}
\author{Z.~Barda\v{c}ov\'{a}}
\affiliation{Comenius University, Bratislava, 81499 Slovakia}
\affiliation{Czech Technical University in Prague, Institute of Experimental and Applied Physics, 11000 Prague, Czech Republic}
\author{I.~A.~Belolaptikov}
\affiliation{Joint Institute for Nuclear Research, Dubna, 141980  Russia}
\author{E.~A.~Bondarev}
\affiliation{Institute for Nuclear
Research of the Russian Academy of Sciences, 60th October Anniversary Prospect 7a, Moscow 117312, Russia}
\author{I.~V.~Borina}
\affiliation{Joint Institute for Nuclear Research, Dubna, 141980  Russia}
\author{N.~M.~Budnev}
\affiliation{Irkutsk State University, Irkutsk, 664003 Russia}
\author{V.~A.~Chadymov}
\affiliation{Independent researcher}
\author{A.~S.~Chepurnov}
\affiliation{Skobeltsyn Research Institute of Nuclear Physics, Lomonosov Moscow State University, Moscow, 119991 Russia}
\author{V.~Y.~Dik}
\affiliation{Joint Institute for Nuclear Research, Dubna, 141980  Russia}
\affiliation{Institute of Nuclear Physics ME RK, Almaty, 050032 Kazakhstan}
\author{G.~V.~Domogatsky}
\altaffiliation{deceased}
\affiliation{Institute for Nuclear
Research of the Russian Academy of Sciences, 60th October Anniversary Prospect 7a, Moscow 117312, Russia}
\author{A.~A.~Doroshenko}
\affiliation{Institute for Nuclear
Research of the Russian Academy of Sciences, 60th October Anniversary Prospect 7a, Moscow 117312, Russia}
\author{R.~Dvornick\'{y}}
\affiliation{Comenius University, Bratislava, 81499 Slovakia}
\author{A.~N.~Dyachok}
\affiliation{Irkutsk State University, Irkutsk, 664003 Russia}
\author{Zh.-A.~M.~Dzhilkibaev}
\affiliation{Institute for Nuclear
Research of the Russian Academy of Sciences, 60th October Anniversary Prospect 7a, Moscow 117312, Russia}
\author{E.~Eckerov\'{a}}
\affiliation{Comenius University, Bratislava, 81499 Slovakia}
\affiliation{Czech Technical University in Prague, Institute of Experimental and Applied Physics, 11000 Prague, Czech Republic}
\author{T.~V.~Elzhov}
\affiliation{Joint Institute for Nuclear Research, Dubna, 141980  Russia}
\author{V.~N.~Fomin}
\affiliation{Independent researcher}
\author{A.~R.~Gafarov}
\affiliation{Irkutsk State University, Irkutsk, 664003 Russia}
\author{K.~V.~Golubkov}
\affiliation{Institute for Nuclear
Research of the Russian Academy of Sciences, 60th October Anniversary Prospect 7a, Moscow 117312, Russia}
\author{N.~S.~Gorshkov}
\affiliation{Joint Institute for Nuclear Research, Dubna, 141980  Russia}
\author{T.~I.~Gress}
\affiliation{Irkutsk State University, Irkutsk, 664003 Russia}
\author{K.~G.~Kebkal}
\affiliation{LATENA, St. Petersburg, 199106, Russia}
\author{V.~K.~Kebkal}
\affiliation{LATENA, St. Petersburg, 199106, Russia}
\author{I.~V.~Kharuk}
\affiliation{Institute for Nuclear
Research of the Russian Academy of Sciences, 60th October Anniversary Prospect 7a, Moscow 117312, Russia}
\author{E.~V.~Khramov}
\affiliation{Joint Institute for Nuclear Research, Dubna, 141980  Russia}
\author{M.~I.~Kleimenov}
\affiliation{Institute for Nuclear
Research of the Russian Academy of Sciences, 60th October Anniversary Prospect 7a, Moscow 117312, Russia}
\author{M.~M.~Kolbin}
\affiliation{Joint Institute for Nuclear Research, Dubna, 141980  Russia}
\author{S.~O.~Koligaev}
\affiliation{INFRAD, Dubna, 141981, Russia}
\author{K.~V.~Konischev}
\affiliation{Joint Institute for Nuclear Research, Dubna, 141980  Russia}
\author{A.~V.~Korobchenko}
\affiliation{Joint Institute for Nuclear Research, Dubna, 141980  Russia}
\author{A.~P.~Koshechkin}
\affiliation{Institute for Nuclear
Research of the Russian Academy of Sciences, 60th October Anniversary Prospect 7a, Moscow 117312, Russia}
\author{V.~A.~Kozhin}
\affiliation{Skobeltsyn Research Institute of Nuclear Physics, Lomonosov Moscow State University, Moscow, 119991 Russia}
\author{M.~V.~Kruglov}
\affiliation{Joint Institute for Nuclear Research, Dubna, 141980  Russia}
\author{V.~F.~Kulepov}
\affiliation{Nizhny Novgorod State Technical University, Nizhny Novgorod, 603950 Russia}
\author{A.~A.~Kulikov}
\affiliation{Irkutsk State University, Irkutsk, 664003 Russia}
\author{Y.~E.~Lemeshev}
\affiliation{Irkutsk State University, Irkutsk, 664003 Russia}
\author{R.~R.~Mirgazov}
\affiliation{Irkutsk State University, Irkutsk, 664003 Russia}
\author{D.~V.~Naumov}
\affiliation{Joint Institute for Nuclear Research, Dubna, 141980  Russia}
\author{A.~S.~Nikolaev}
\affiliation{Skobeltsyn Research Institute of Nuclear Physics, Lomonosov Moscow State University, Moscow, 119991 Russia}
\author{I.~A.~Perevalova}
\affiliation{Irkutsk State University, Irkutsk, 664003 Russia}
\author{D.~P.~Petukhov}
\affiliation{Institute for Nuclear
Research of the Russian Academy of Sciences, 60th October Anniversary Prospect 7a, Moscow 117312, Russia}
\author{E.~N.~Pliskovsky}
\affiliation{Joint Institute for Nuclear Research, Dubna, 141980  Russia}
\author{M.~I.~Rozanov}
\affiliation{Sankt Petersburg State Marine Technical University, Sankt Petersburg, 190008 Russia}
\author{E.~V.~Ryabov}
\affiliation{Irkutsk State University, Irkutsk, 664003 Russia}
\author{G.~B.~Safronov}
\affiliation{Institute for Nuclear
Research of the Russian Academy of Sciences, 60th October Anniversary Prospect 7a, Moscow 117312, Russia}
\author{B.~A.~Shaybonov}
\affiliation{Joint Institute for Nuclear Research, Dubna, 141980  Russia}
\author{V.~Y.~Shishkin}
\affiliation{Skobeltsyn Research Institute of Nuclear Physics, Lomonosov Moscow State University, Moscow, 119991 Russia}
\author{E.~V.~Shirokov}
\affiliation{Skobeltsyn Research Institute of Nuclear Physics, Lomonosov Moscow State University, Moscow, 119991 Russia}
\author{F.~\v{S}imkovic}
\affiliation{Comenius University, Bratislava, 81499 Slovakia}
\affiliation{Czech Technical University in Prague, Institute of Experimental and Applied Physics, 11000 Prague, Czech Republic}
\author{A.~E. Sirenko}
\affiliation{Joint Institute for Nuclear Research, Dubna, 141980  Russia}
\author{A.~V.~Skurikhin}
\affiliation{Skobeltsyn Research Institute of Nuclear Physics, Lomonosov Moscow State University, Moscow, 119991 Russia}
\author{A.~G.~Solovjev}
\affiliation{Joint Institute for Nuclear Research, Dubna, 141980  Russia}
\author{M.~N.~Sorokovikov}
\affiliation{Joint Institute for Nuclear Research, Dubna, 141980  Russia}
\author{I.~\v{S}tekl}
\affiliation{Czech Technical University in Prague, Institute of Experimental and Applied Physics, 11000 Prague, Czech Republic}
\author{A.~P.~Stromakov}
\affiliation{Institute for Nuclear
Research of the Russian Academy of Sciences, 60th October Anniversary Prospect 7a, Moscow 117312, Russia}
\author{O.V.~Suvorova}
\affiliation{Institute for Nuclear
Research of the Russian Academy of Sciences, 60th October Anniversary Prospect 7a, Moscow 117312, Russia}
\author{V.~A.~Tabolenko}
\affiliation{Irkutsk State University, Irkutsk, 664003 Russia}
\author{V.~I.~Tretjak}
\affiliation{Joint Institute for Nuclear Research, Dubna, 141980  Russia}
\author{B.~B.~Ulzutuev}
\affiliation{Joint Institute for Nuclear Research, Dubna, 141980  Russia}
\author{Y.~V.~Yablokova}
\affiliation{Joint Institute for Nuclear Research, Dubna, 141980  Russia}
\author{D.~N.~Zaborov}
\affiliation{Institute for Nuclear
Research of the Russian Academy of Sciences, 60th October Anniversary Prospect 7a, Moscow 117312, Russia}
\author{S.~I.~Zavyalov}
\affiliation{Joint Institute for Nuclear Research, Dubna, 141980  Russia}
\author{D.~Y.~Zvezdov}
\affiliation{Joint Institute for Nuclear Research, Dubna, 141980  Russia}
\collaboration{66}{(Baikal-GVD Collaboration)}  
\author{Y.~Y.~Kovalev}
\affiliation{Max-Planck-Institut f\"ur Radioastronomie, Auf dem H\"ugel 69, 53121 Bonn, Germany}
\author{A.~V.~Plavin}
\affiliation{Black Hole Initiative at Harvard University, 20 Garden Street, Cambridge, MA 02138, USA}
\author{D.~V.~Semikoz}
\affiliation{APC, Universit\'e Paris Diderot, CNRS/IN2P3, CEA/IRFU, Sorbonne Paris Cit\'e, 119 75205 Paris, France}
\author{S.~V.~Troitsky}
\affiliation{Institute for Nuclear
Research of the Russian Academy of Sciences, 60th October Anniversary Prospect 7a, Moscow 117312, Russia}
\affiliation{Physics Department, Lomonosov Moscow State University, 1-2 Leninskie Gory, Moscow 119991, Russia}

\begin{abstract}
Recent observations of the Galactic component of the high-energy neutrino flux, together with the detection of the diffuse Galactic gamma-ray emission up to sub-PeV energies, open new possibilities to study the acceleration and propagation of cosmic rays in the Milky Way. At the same time, both large non-astrophysical backgrounds at TeV energies and scarcity of neutrino events in the sub-PeV band currently limit these analyses. Here we use the sample of cascade events with estimated neutrino energies above 200~TeV, detected by the partially deployed Baikal Gigaton Volume Detector (GVD) in six years of operation, to test the continuation of the Galactic neutrino spectrum to sub-PeV energies. We find that the distribution of the arrival directions of Baikal-GVD cascades above 200~TeV in the sky suggests an excess of neutrinos from low Galactic latitudes {\red with the chance probability of $1.4 \cdot 10^{-2}$.}
We find the excess above 200 TeV also in the most recent IceCube public data sets, both of cascades and tracks. {\red The chance probability of the excess 
in the combined IceCube and Baikal-GVD analysis is $3.4 \cdot 10^{-4}$.  }
The flux \black of Galactic neutrinos above 200~TeV challenges often-used templates for neutrino search based on cosmic-ray simulations.
\end{abstract}



\section{Introduction}
\label{sec:Introduction}
The origin of cosmic rays with energies between $\sim 10^{12}$~eV and $\sim 10^{20}$~eV was puzzling for decades. The observation of high-energy astrophysical neutrinos by the IceCube experiment \citep{IceCube-2013,IceCube-HESE-2020}, recently confirmed by the Baikal Gigaton Volume Detector \citep[GVD;][]{Baikal-diffuse}, has opened a new view on this old question. Indeed, these neutrinos are most probably born, together with photons, in interactions of energetic cosmic rays with matter and radiation. Unlike charged cosmic rays, neutrinos are not deflected by cosmic magnetic fields and thus point back to the place where they are produced. Unlike photons, neutrinos are not absorbed or scattered and thus reach the observer from distant or opaque sources. Despite complications related to the large atmospheric background and to relatively low precision of the reconstruction of individual events, high-energy neutrino astronomy has developed into an important new branch of astrophysics, see e.g.\ \citet{ST-UFN,ST-UFN1} for reviews.

Of particular interest is the neutrino radiation coming from our Galaxy, which is expected \citep{2comp-Chen,2comp-Vissani,2comp-Neronov,2comp-Vissani-2} to supplement the extragalactic contribution. Despite numerous early attempts \citep{Neronov:2013lza,2016APh....75...60N,ST-Gal,Denton,IceCubeANTARES-GalPlane,IceCube:cascades-Gal2sigma}, the existence of the Galactic neutrino flux has been established only recently, in three independent data sets \citep{neutgalaxy,ANTARES-ridge,IceCube-Galaxy}. The three results, obtained with different techniques and testing different parts of the Milky Way, demonstrate overall order-of-magnitude consistency between each other, as well as with the inference from observations of diffuse gamma rays by Tibet-AS$\gamma$ \citep{Tibet-GalDiffuse} and LHAASO \citep{LHAASO-GalDiffuse}, see e.g.\ Fig.~5 of \citet{ST-UFN1} and discussion there. However, considerable differences in best-fit normalizations are present even between different templates used by \citet{IceCube-Galaxy} for the search of the Galactic-plane signal with IceCube cascade events. Moreover, a model-independent analysis of published IceCube tracks demonstrates \citep{neutgalaxy} a significant Galactic excess at neutrino energies above 200~TeV, which does not match predictions of the templates both in the spectrum and in the spatial distribution of the signal. Studies of these tensions open up the possibility to improve contemporary models of the Galactic cosmic rays.

With the current instrumented volume of $\sim 0.6$~km$^3$ (and growing), and having better angular resolution thanks to the liquid water with respect to ice, Baikal-GVD is properly suited for studies of the Galactic neutrino signal at the highest neutrino energies. Here we report on the observation of the Milky Way with Baikal-GVD cascade events above 200~TeV, consistent with \citet{neutgalaxy}. We also consider new publicly available sets of IceCube cascade and track events above 200~TeV and find that the Galactic signal in these data is consistent with our results.

In Sec.~\ref{sec:data}, we briefly describe Baikal-GVD and its updated cascade data set. Section~\ref{sec:tests} describes the analysis of the data set and its results. In Sec.~\ref{sec:disc}, we compare the Baikal-GVD Milky-Way result with those obtained from IceCube data, and discuss astrophysical implications of our observation. Section~\ref{sec:concl} presents our brief conclusions.

\section{Data}
\label{sec:data}
Baikal-GVD is the largest neutrino telescope currently operating in the Northern hemisphere (latitude $51.5^\circ$ N). Like other water Cerenkov instruments, it may detect neutrino-induced events as cascades and tracks, with very different sensitivities and analysis procedures. Details of the experiment, event selection and analysis can be found e.g.\ in \cite{Baikal-JETP,Baikal-diffuse,Baikal-tracks,Baikal-TXS}, and we do not repeat them here.

The high-energy cascade sample was described by \citet{Baikal-diffuse}. It contains events with reconstructed energies $E\ge 70$~TeV and the expected probability of their astrophysical origin $>50\%$, estimated from simulations. Compared to \citet{Baikal-diffuse}, we add two more years of data collection. The telescope consists of clusters of optical modules, currently 13, with each cluster operated as an independent unit. Since the telescope is growing, with new clusters added every spring, these two years almost doubled the exposure, which corresponds to $\approx 26.8$ years of one-cluster operation in Spring 2018 -- Spring 2024.

Following the previous study \citep{neutgalaxy}, we consider events with $E\ge 200$~TeV. 
Table~\ref{tab:events}
\begin{table}[t]
\caption{\label{tab:events} List of Baikal-GVD cascades with reconstructed neutrino energies $E\ge 200$~TeV, observed in 2018--2023 observational seasons. Presented are energies $E$, Galactic coordinates $(l,b)$, 50\% CL and 90\% CL accuracies of the determination of the arrival direction, $r_{50}$ and $r_{90}$, respectively.}
\centering
\begin{tabular}{cccccc}
\hline\hline
Event ID & $E$, TeV & $l$,$^\circ$ & $b$,$^\circ$ & $r_{50}$,$^\circ$ & $r_{90}$,$^\circ$\\
\hline
 \text{GVD190517CA} & 1200 & 99.9 & 54.9 & 2.0 & 3.0 \\
 \text{GVD210117CA} & 246 & 168.8 & 38.8 & 1.6 & 3.6 \\
 \text{GVD210409CA} & 263 & 73.3 & $-$6.1 & 3.3 & 6.3 \\
 \text{GVD210418CA} & 224 & 196.8 & $-$14.6 & 3.0 & 5.8 \\
 \text{GVD221112CA} & 380 & 61.0 & $-$4.7 & 2.9 & 7.7 \\
 \text{GVD230518CA} & 214 & 199.0 & 4.7 & 2.3 & 4.7 \\
 \text{GVD231006CA} & 245 & 76.9 & 5.3 & 2.3 & 5.1 \\
 \text{GVD230611CA} & 479 & 15.2 & 36.2 & 2.6 & 5.2 \\
\hline\hline
\end{tabular}
\end{table}
presents the list of 8 events used in this analysis. \red In agreement with simulations, \black only one of these events comes from below the horizon, because the Earth becomes opaque for neutrinos of sub-PeV energies. Monte-Carlo simulations indicate that about $64\%$ of the events passing the selection criteria and having $E\ge 200$~TeV are expected to have the astrophysical origin.

\section{Search for Galactic neutrinos}
\label{sec:tests}
In the present study, we adopt the model-independent approach used by \citet{neutgalaxy}. It does not rely on any assumptions about the origin and properties of the Galactic signal and tests only the excess of events from the Galactic plane. We introduce a single nonparametric test statistics, median of the absolute value of the Galactic latitude, $|b|_{\rm med}$, calculated over the events sample. Following \citet{neutgalaxy}, we use the events with best-fit reconstructed energies $E \ge 200$~TeV. The second selection cut of \citet{neutgalaxy}, the area of the track direction error region in the sky, is irrelevant for cascade events studied here. Therefore, the present study represents a direct test of the observation of \citet{neutgalaxy} with completely independent data.

To search for the possible excess of events from the Galactic plane, which would decrease $|b|_{\rm med}$, we compare the observed value of $|b|_{\rm med}$ with that expected for a distribution of arrival directions having no Galactic excess. This distribution is not isotropic because of contributions of both atmospheric and extragalactic events, and further correction because of the non-uniform energy-dependent experimental exposure. However, for a continuously operating installation, like Baikal-GVD, the Earth's rotation makes this distribution independent from the Right Ascension (RA). Therefore, reshuffling RA values of observed events provides for a robust data-driven way to generate random sets of
arrival directions, used multiple times in the analysis of data of various neutrino telescopes, see e.g.\ \citet{scramble-Baikal,scramble-IceCube,scramble-ANTARES}. In this way, we generate $10^{5}$ artificial sets of 8 events each and calculate $|b|_{\rm med}$ for each of them.

\begin{figure}
\centering
\includegraphics[width=0.98\columnwidth]{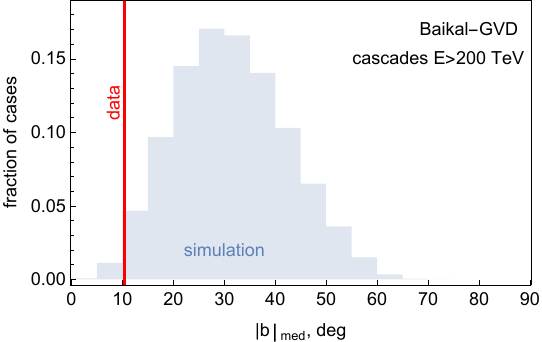}
\caption{Distribution (shaded histogram) of the median $|b|_{\rm med}$ in simulated sets of Baikal-GVD cascades with $E\ge 200$~TeV. The observed value of $|b|_{\rm med}$ is shown by the vertical red line.
\label{fig:bmed-hist-GVD}
}
\end{figure}

For the real data set, $|b|_{\rm med}= 10.4^\circ$, while the value expected from simulations is $\langle |b|_{\rm med} \rangle = 31.4^\circ$, which indicates the presence of the Galactic excess in the data, see Fig.~\ref{fig:bmed-hist-GVD}.
To assess the significance of the excess, we estimate the fraction of realizations of simulated data sets for which the value of $|b|_{\rm med}$ does not exceed the observed one. This gives the p-value of the rejection of the hypothesis of the absence of the Galactic excess, $p=1.4 \times 10^{-2}$, see Fig.~\ref{fig:bmed-hist-GVD}. Note that the study does not have any trials, therefore this value should be treated as the post-trial one. 
\red It is customary to illustrate the rejection p-values with corresponding significances for two-sided Gaussian distribution. Hereafter we quote these significances, keeping in mind that only p-values are meaningful for non-Gaussian statistics. The rejection of the absence of the Galactic excess with Baikal-GVD cascades would correspond to $2.5\sigma$ in this interpretation. \black

\begin{figure}
\centering
\includegraphics[width=0.98\columnwidth]{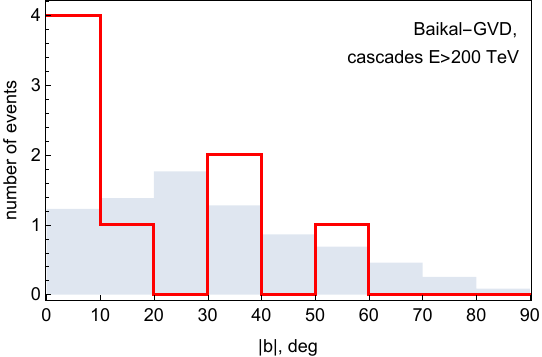}
\caption{Observed (red line) and expected (shaded histogram) distribution of $|b|$ for Baikal-GVD cascades with $E\ge 200$~TeV.
\label{fig:bdist-hist-GVD}
}
\end{figure}

Given the size of the event sample, it would be difficult to measure the spectrum, and even the normalization, of the Galactic neutrino flux. For a very rough estimate, we examine the distribution of observed and simulated events in $|b|$, see Fig.~\ref{fig:bdist-hist-GVD}, 
and make use of the Poisson distribution to find the excess number of events with $|b|<10^{\circ}$ to be \red $n_{\rm MW}=2.8^{+3.2}_{-1.2}$. \black We compare it with the total expected number of astrophysical events in the sample, $n_{\rm astro}=5.1$, and estimate the Galactic neutrino flux as the fraction of the total full-sky astrophysical neutrino flux measured by Baikal-GVD with cascades
\citep{Baikal-diffuse},
$$
F_{\rm MW}=  \frac{n_{\rm MW}}{n_{\rm astro}} \red \xi \black  F_{\rm astro} =
4\pi 10^{-18}\phi_{\rm MW} \left( \frac{E}{100~\mbox{TeV}} \right)^{-\gamma},
$$
with $\phi_{\rm MW}=\red 1.6^{+2.0}_{-0.9}~ \black \mbox{GeV}^{-1} \mbox{cm}^{-2} \mbox{s}^{-1} $ and $\gamma=2.58^{+0.27}_{-0.33}$
(like in \cite{Baikal-diffuse}, this is the total flux of neutrinos and antineutrinos per flavor, assuming flavor equipartition).
\red
Here we introduced the coefficient $\xi$ related to the difference in the exposures for $|b|<10^\circ$ and isotropic astrophysical neutrinos, which have different distributions in the zenith angles. Unlike for the main analysis in terms of $|b|_{\rm med}$, here we use Monte-Carlo simulations \citep{Baikal-diffuse} of the atmospheric and astrophysical neutrinos to determine both $\xi$ and the expected distribution in $|b|$ of non-Galactic events. We have verified that reshuffling RA of real events gives quantitatively similar results, which are consistent with the MC-based estimates within the statistical uncertainties due to the limited number of events available for reshuffling. \black

\section{Discussion}
\label{sec:disc}
\subsection{Comparison with IceCube data  at $E\ge 200$~TeV}
\label{sec:disc:IceCube}

\citet{neutgalaxy} studied a compilation of publicly available data on IceCube tracks with estimated energies $E\ge 200$~TeV and found a statistically significant excess from events from low Galactic latitudes. The present study confirms this result with the Baikal-GVD data. However, new IceCube sets of both cascades and tracks have recently become available for the public, and we use them to search for the Galactic neutrino component above 200~TeV by exactly the same method.

There are 12 high-energy starting cascade events (HESE) with $E\ge 200$~TeV reported by \citet{HESE-data-paper,HESE-data}. Note that the astrophysical purity of the HESE data set at these energies, $\sim 95\%$, is higher than that for the Baikal-GVD set we use here, $\sim 64\%$, because of different selection cuts. At the same time, the total exposure of Baikal-GVD is 20.9~m$^2\cdot$yr for this data set, while that of IceCube HESE sample we use ($E\ge 200$~TeV) may be estimated as 176~m$^2\cdot$yr based on the effective area \citep{IceCube-HESE-2020} and the exposure time of 12~years. Note that the HESE sample includes 4 starting tracks in addition to 12 cascades we use here. The total numbers of events in both sets, 16 in HESE (expected 22.8) and 8 in Baikal-GVD cascades (expected 8.3), agrees with the experiments' exposures at 5\%~CL and 45\%~CL, respectively\footnote{For the estimates in this paragraph, the best-fit power-law fluxes obtained from the HESE \citep{IceCube-HESE-2020} and Baikal-GVD \citep{Baikal-diffuse} samples are used.}.  

Applying the procedure described in Sec.~\ref{sec:tests} to the IceCube HESE data set, we find a similar Galactic excess because the observed $|b|_{\rm med}= 12.4^\circ$, while the expected $\langle |b|_{\rm med} \rangle = 31.9^\circ $. The p-value for this excess $p=8.7 \cdot 10^{-3}$ ($2.6\sigma$).

The recent public uniform compilation of high-energy IceCube track events is presented in the ICECAT catalog \citep{ICECAT-paper,ICECAT-data}, recently updated to its version~2. Making use of the same cuts defined by \citet{neutgalaxy}, that is requiring the best-fit $E\ge 200$~TeV and the 90\% CL area of uncertainty in the track direction below 10 square degrees, we are left with 67 events, with the average astrophysical purity of this sample $\sim 65\%$. This sample is not independent from that of \cite{neutgalaxy}, having a considerable overlap, though the energies and directions reported in ICECAT were obtained with a different reconstruction procedure. Not surprisingly, this sample also demonstrates the Galactic excess, with the observed $|b|_{\rm med}= 24.7^\circ$, expected $\langle |b|_{\rm med} \rangle =  36.0^\circ$, and $p=1.8 \cdot 10^{-3}$ ($3.1\sigma$).

\begin{figure}[t]
\centering
\includegraphics[width=0.98\columnwidth]{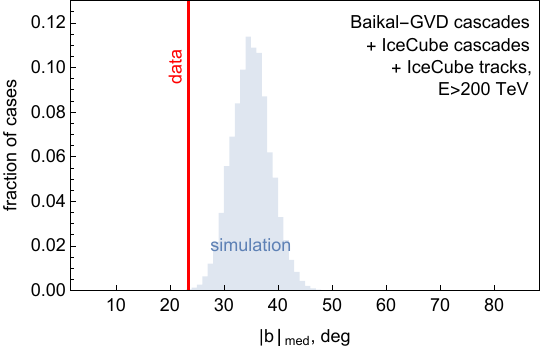}
\caption{Distribution (shaded histogram) of the median $|b|_{\rm med}$ in simulated combined sets of Baikal-GVD cascades, IceCube cascades and IceCube tracks with $E\ge 200$~TeV. The observed value of $|b|_{\rm med}$ is shown by the vertical red line.
\label{fig:bmed-hist-all3}
}
\end{figure}

We see that all three data sets, Baikal-GVD cascades, IceCube cascades, and IceCube tracks (all above 200~TeV), demonstrate the excess of events close to the Galactic plane, also visible in the sky map, Fig.~\ref{fig:map}. We also perform a combined analysis of all three samples in the same manner, resulting in $p=3.4 \cdot 10^{-4}$ ($3.6\sigma$), see Fig.~\ref{fig:bmed-hist-all3}. 
For convenience, we collect the results of the three analyses performed here, and of their combination, in Table~\ref{tab:bmed}.
\begin{table}
\caption{Results (this work) of the search for the Galactic component of the neutrino flux above 200~TeV (see the text for details).}
\label{tab:bmed}
\centering
\begin{tabular}{cccc}
\hline
\hline
Sample 
 & $|b|_{\rm med}$ & $\langle|b|_{\rm med}\rangle$ &  $p$   \\
 & observed & expected  &   \\
\hline
cascades: &&&\\
Baikal-GVD &10.4$^\circ$ & 31.4$^\circ$ & $1.4 \cdot 10^{-2}$  ($2.5\sigma$) \\
IceCube  &12.4$^\circ$ & 31.9$^\circ$ & $8.7 \cdot 10^{-3}$  ($2.6\sigma$) \\
combined  &12.4$^\circ$ & 31.5$^\circ$ & $1.7 \cdot 10^{-3}$  ($3.1\sigma$) \\
\hline
IceCube tracks             &24.7$^\circ$ & 36.0$^\circ$ & $1.8 \cdot 10^{-3}$  ($3.1\sigma$) \\
\hline
all combined             &23.4$^\circ$ & 35.0$^\circ$ & $3.4 \cdot 10^{-4}$  ($3.6\sigma$) \\
\hline
\end{tabular}
\end{table}

In each set, we estimate the fraction of events from the Milky Way in the total astrophysical neutrino flux above 200~TeV  as described in Sec.~\ref{sec:tests} \red for cascades, $|b|<10^\circ$, \black and in \citet{neutgalaxy} \red for tracks, $|b|<20^\circ$\black. Making use of these fractions and the total fluxes measured in different analyses, we obtain rough estimates of the full-sky Milky-Way neutrino flux at energies between 200~TeV and 1~PeV.
\red Here, \black we use the astrophysical flux of \citet{Baikal-diffuse} for Baikal-GVD cascades, of \citet{IceCube-HESE-2020} for IceCube HESE events, of \citet{IceCube:cascade-spec} for lower-energy cascades, and of \red \citet{IceCube:muon2018} \black for IceCube tracks\red\footnote{\red Energies of ICECAT events were estimated \citep{ICECAT-data} assuming this older spectrum.}. 
The flux estimates obtained in this way are collected \black in Table~\ref{tab:fluxes}.
\begin{table}
\caption{Integral fluxes (in units of $10^{-13}$~cm$^{-2}$s$^{-1}$) of Galactic neutrinos with $200\,\mathrm{TeV}<E<1\,\mathrm{PeV}$ and the Galactic fractions in the total astrophysical flux (per flavor) of neutrinos at these energies, obtained in different analyses.}
\label{tab:fluxes}
\centering
\begin{tabular}{ccc}
\hline
\hline
    & Flux & Fraction \\
\hline
\multicolumn{3}{c}{Predicted by templates:}\\
\hline
KRA$\gamma_5$    & 0.34   &  --  \\
KRA$\gamma_{50}$ & 0.78   &  --  \\
$\pi^0$          & 0.077   &  --  \\
\hline
\multicolumn{3}{c}{Templates normalized to IceCube \citep{IceCube-Galaxy}:}\\
\hline
KRA$\gamma_5$    & $0.19^{+0.06}_{-0.05}$   & $0.044^{+0.016}_{-0.014}$  \\
KRA$\gamma_{50}$ & $0.29^{+0.10}_{-0.09}$   & $0.067^{+0.026}_{-0.024}$  \\
$\pi^0$          & $0.37^{+0.09}_{-0.08}$   & $0.086^{+0.026}_{-0.025}$   \\
\hline
\multicolumn{3}{c}{Estimated by \citet{neutgalaxy}:}\\
\hline
IceCube tracks     & $1.3\pm 0.5$   & $0.28\pm 0.09$  \\
\hline
\multicolumn{3}{c}{Estimated in the present work:}\\
\hline
Baikal-GVD cascades&\red $3.9^{+5.0}_{-2.7}$   &\red $0.52^{+0.60}_{-0.21}$  \\
IceCube cascades   & $1.0^{+1.2}_{-0.6}$   & $0.26^{+0.30}_{-0.12}$  \\
IceCube tracks     &\red $0.9^{+0.7}_{-0.5}$   & \red $0.22^{+0.15}_{-0.10}$  \\
\hline
\end{tabular}
\end{table}
One can see that our results for Baikal-GVD cascades, IceCube cascades and IceCube tracks are in a good agreement, given the uncertainties. 
\red 
Note that the statistical uncertainties in these flux estimates are large because of the low number of events associated with the Milky Way. We consider the results of the model-independent $|b_{\rm med}|$ test, see Table~\ref{tab:bmed}, as the main results of our study.
\black

\subsection{Implications}
\label{sec:disc:implications}

In Table~\ref{tab:fluxes}, we also present the Galactic neutrino fluxes between 200~TeV and 1~PeV predicted in three spectral templates \citep{KRAgamma,pi0} assumed in IceCube studies \citep{IceCube-Galaxy}. The normalizations of the template spectra have been kept free by \citet{IceCube-Galaxy}, and, in addition, we estimate the Galactic fluxes and fractions for the best-fit normalizations of the three templates. One can see a dramatic difference between template predictions and our model-independent results above 200~TeV: previously used spectral templates underpredict the Galactic neutrino flux at these energies.

One of the best motivated mechanism for the production of Galactic neutrinos assumes interaction of energetic cosmic rays with ambient matter, which are saturated by pi-meson production. While decays of charged $\pi^{\pm}$ give birth to the neutrinos, their neutral counterparts $\pi^{0}$ decay to energetic photons, so the fluxes of the two messengers become related, see e.g.\ \citet{ST-UFN} and references therein. Unlike from extragalactic sources, these photons reach us from the Milky Way with modest to no attenuation. Diffuse fluxes of such very energetic Galactic gamma rays have been observed by Tibet-AS$\gamma$ \citep{Tibet-GalDiffuse} and LHAASO \citep{LHAASO-GalDiffuse} experiments. 

\begin{figure}
\centering
\includegraphics[width=0.98\columnwidth]{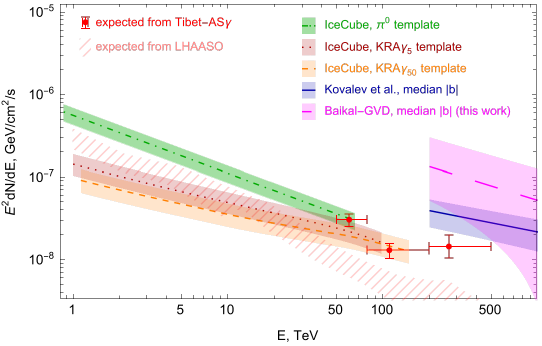}
\caption{Estimated full-sky spectra of Galactic neutrinos \red (per one flavor of neutrino plus antineutrino) \black obtained in the present and in some of preceding studies, together with those expected from observations of diffuse Galactic gamma rays. See the plot legend for notations and \cite{ST-UFN1} for details and further references. 
\label{fig:specplot}
}
\end{figure}

\begin{figure*}
\centering
\includegraphics[width=0.85\linewidth]{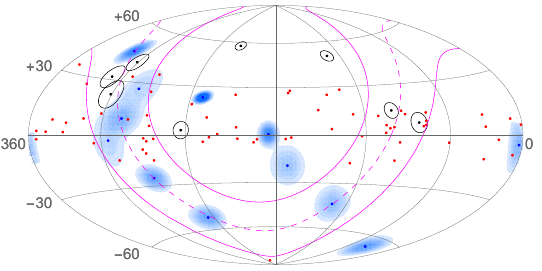}
\caption{Arrival directions of Baikal-GVD (black projected circles of $r_{90}$ radius) and IceCube (shading presenting the likelihood of the direction) cascade events, as well as of IceCube track events (red dots) with $E\ge 200$~TeV in the sky map in equatorial coordinates. The dashed magenta line represents the Galactic plane, and two full magenta lines limit the zone $|b|<20^\circ$.
\label{fig:map}
}
\end{figure*}

Figure~\ref{fig:specplot}
presents the observed Galactic neutrino flux from Baikal-GVD cascades, estimated in the present work, together with expectations from Tibet-AS$\gamma$ and LHAASO observations (see \citet{FangMurase2023,ST-UFN1} for details), in the assumption of the common origin of both neutrinos and photons in the proton collisions. The difference between two experiments in the gamma-ray fluxes at high energies may be related to different masks imposed to cut point sources of high-energy emission. In this case, the fact that the Milky-Way neutrino emission fits better the expectations from Tibet-AS$\gamma$ than those from LHAASO, might indicate that the Galactic neutrino emission above 200~TeV comes, at least partially, from individual sources, Galactic PeVatrons. Indeed, the sky map of the neutrino events studied here, Fig.~\ref{fig:map},
suggests some clustering of cascade events towards the Cygnus region, which also manifests itself in gamma rays \citep{Tibet-GalDiffuse}. 
Moreover, recently LHAASO detected significant gamma-ray flux in a $\sim 6^\circ$ size halo Cygnus region with gamma-rays up to to PeV energies distributed across this region \citep{LHAASO:2023uhj}. ICECAT has very low exposure towards this region at high energies, so it is hardly possible to test this concentration with IceCube tracks.

\section{Conclusions}
\label{sec:concl}

By analyzing cascade events with estimated neutrino energies above 200~TeV, observed by Baikal-GVD during six years of operation, we find the concentration of events towards the Galactic plane, indicating the presence of a large Galactic component in the high-energy astrophysical neutrino flux, with the p-value of the absence of the Galactic component of $p=1.4 \cdot 10^{-2}$ obtained in a nonparametric, model-independent approach. The estimated Galactic neutrino flux above 200~TeV matches the one obtained by \cite{neutgalaxy} for IceCube tracks in the same energy range. We test that the similar results hold for the most recent publicly available IceCube samples of both cascades and tracks, with the p-value of $3.4\cdot 10^{-4}$ obtained in the combined analysis of the three samples by the same method.

The Galactic neutrino flux agrees with the expectations from the gamma-ray diffuse Milky-Way emission observed by Tibet-AS$\gamma$, though a direct comparison requires model-dependent assumptions. The neutrino flux is somewhat higher than similar expectations from LHAASO observations. This may indicate that the neutrino emission is not purely diffuse, and some part of it comes from localized, point-like or extended, sources, masked in the LHAASO analysis. The Cygnus region, seen in the neutrino sky map, may host some of them \red\citep{BykovCygnusPSR,IceCube-binaries,Semikoz_Cygnus,Cygnus2024}. \black

The Galactic neutrino component at very high energies is so prominent that is clearly detected despite low statistics. The fraction of Galactic events in the total astrophysical flux above 200~TeV reaches several tens per cent, which is in a disagreement with assumptions of many model-dependent analyses, including that of \citet{IceCube-Galaxy}. Together with the distribution of observed arrival directions in the sky, which suggested \citep{neutgalaxy} a wider Milky Way in neutrinos than predicted by models, this observation challenges contemporary scenarios of cosmic-ray acceleration and propagation in the Galaxy. As it has been previously pointed out by \citet{neutgalaxy} and \citet{ST-UFN1}, explaining this shape may require significant contribution of neutrinos from the local origin to the total flux, cf.\ \citet{LocalBubble1,LocalBubble2,GiacSemikoz}.

Baikal-GVD continues to collect data, gradually increasing its instrumented volume. The upcoming Baikal-GVD data sets, including track-like events, as well as data from the other neutrino telescopes, promise exciting prospects to test the intriguing observations presented here.

\medskip
This work is supported in the framework of the State project ``Science'' by the Ministry of Science and Higher Education of the Russian Federation under the contract 075-15-2024-541.
The work of A.P.\ was supported by the Black Hole Initiative, which is funded by grants from the John Templeton Foundation (Grant \#60477, 61479, 62286) and the Gordon and Betty Moore Foundation (Grant GBMF-8273).

\bibliography{GVD-Galactic}

\end{document}